\documentclass[useAMS,usenatbib,usegraphicx]{mn2e}

\newcommand{\BV}{Brunt-V\"ais\"al\"a }
\newcommand{\teff}{\mbox{${T}_{\rm eff}$}}

\newcommand{\rsol}{\mbox{${\rm R}_{\odot}$}}
\newcommand{\msol}{\mbox{${\rm M}_{\odot}$}}
\newcommand{\msun}{\mbox{${\rm M}_{\odot}$}}
\newcommand{\lsol}{\mbox{${\rm L}_{\odot}$}}

\newcommand{\simgt}{\lower.5ex\hbox{$\; \buildrel > \over \sim \;$}}
\newcommand{\simlt}{\lower.5ex\hbox{$\; \buildrel < \over \sim \;$}}

\title[12 Bo\"otis: a test bed for extra-mixing processes in stars]{12 Bo\"otis: a test bed for extra-mixing processes in stars}
\author[A. Miglio, J. Montalb\'an and C. Maceroni]{Andrea Miglio$^{1}$, Josefina Montalb\'an$^{1}$ and Carla Maceroni$^{2}$\\
$^{1}$Institut d'Astrophysique et de G\'eophysique de l'Universit\'e de Li\`ege,
All\'ee du 6 Ao\^ut, 17 B-4000 Li\`ege, Belgium\\
$^{2}$INAF-Osservatorio di Roma, via Frascati-33, Monteporzio Catone (RM), Italy}

\begin{document}

\date{Accepted 1988 December 15. Received 1988 December 14; in original form 1988 October 11}

\pagerange{\pageref{firstpage}--\pageref{lastpage}} \pubyear{2002}

\maketitle

\label{firstpage}

\begin{abstract}
12 Bo\"otis is a spectroscopic binary whose visual orbit has been resolved by interferometry. Though the physical 
parameters of the system have been determined with an excellent precision, the theoretical modelling of the components is still uncertain.
We study the capability of solar-like oscillations to distinguish between calibrated models of the system obtained by 
including  in the stellar modelling different mixing processes.
We consider different scenarios for the chemical transport processes: classical overshooting, microscopic diffusion 
and turbulent mixing. For each of them we calibrate the stellar models of 12 Boo~A and B by fitting the available 
observational constraints by means of a Levenberg-Marquardt minimization algorithm, and  finally, we analyze the 
asteroseismic properties of different calibrated models.
Several solutions with 12~Boo~A in (or close to) post-main sequence and 12 Boo~B on main sequence are found by assuming 
a thickness of the overshooting layer between 0.06 and 0.23 the pressure scale height. Solutions with both components on the 
main sequence can be found only by assuming an overshoot larger in the primary than in the secondary, or a more efficient 
central mixing for 12 Boo~A than for 12 Boo~B. 
We show that the detection of solar-like oscillations expected in these stars would allow to distinguish  between different scenarios and  provide therefore an estimation of the overshooting parameters and of the properties of extra-mixing processes.
\end{abstract}


\section{Introduction}

  In standard stellar modelling  convection  is described by means of a local theory such as the Mixing Length Theory
 \citep{Bohm58}, and the extension of the convective regions is determined by
the classical Schwarschild criterion ($\nabla_{\rm rad}=\nabla_{\rm ad}$). This is one of the well known shortcomings
of stellar structure and evolution theory. For intermediate and high mass stars, where a convective core is developed, 
the mass  of this central mixed region plays a   fundamental role in determining  the lifetime and the luminosity 
of the main sequence phase, and affects   other relevant  aspects such as the s-process nucleosynthesis on the asymptotic 
giant branch \citep{Ventura05},  the chemical evolution of the interstellar medium, and the size and composition of 
white dwarfs \citep[see, e.g.][]{Chiosi98}.

We mean  here by overshooting or extra-mixed region the  mixing of material beyond the formal boundary of 
convection region set by the Schwarschild criterion. There is quite a large number of observational evidences indicating 
that this phenomenon of overshooting  exists. For instance,  comparisons  between observations \citep[binaries and open 
clusters,][]{Andersen90,Ribas00} and theoretical models indicate that stars have larger convective cores than predicted
by theory. Moreover, numerical simulations and laboratory experiments suggest  the overshoot of convective elements 
in the adjacent stable regions.  A great effort has been done during the last 30 years to describe turbulent convection 
in  stellar interiors modelling, however,  given the difficulty in estimating the rate of dissipation of turbulent kinetic 
energy, it is still not possible to predict the thickness of the extra-mixed region from first principles. Therefore, the
thickness of the overshooted region ($\Lambda_{\rm OV}$)  is usually parameterized in terms of the local pressure scale
height $H_{\rm p}$ ($\Lambda_{\rm OV}=\alpha_{\rm OV} H_{\rm p}$). The value of this parameter has been estimated by  
fitting theoretical isochrones to the observed color-magnitude diagrams of open clusters, and  by modelling  double-lined 
eclipsing binaries with well determined masses and radii \citep[see e.g.][]{Ribas00}. These  calibrations also indicate
that the exact value of the overshoot parameter might depend on the stellar mass: values  of the order of 0-0.1 are 
required for masses between 1.1 and 1.5 M$_{\odot}$, and  values from 0.2 to 0.5   are needed in the 1.6 to 9 M$_{\odot}$ 
interval \citep[see][]{Ribas00}.  Moreover, as noticed in  \citet{Vandenberg04} and \citet{Vandenberg06}, the 
overshooting parameter has probably a steep dependence on mass in the range $1.3 \la M/M_\odot \la 1.55$.
 
Not only the exact amount of overshooting and  its dependence on the stellar mass, but also the physical process
responsible for extra-mixing, are still matter of a lively debate \citep{Maeder98, Young05, Popielski05}.
 
In this context the binary system 12~Bo\"otis represents a valuable observational test of stellar modelling. Recent 
interferometric measurements allowed \cite{Boden05} to determine the masses of the components of this double-lined 
spectroscopic binary with a precision of about 0.3\%.  The masses of the components are very similar (mass ratio 
$q\sim0.97$), but their luminosities are quite different. A preliminary determination of stellar parameters by comparison 
with theoretical isochrones \citep[see][]{Boden05},  suggested that secondary component is still in the central-hydrogen 
burning phase, while the primary is more evolved and burning hydrogen in a shell.  Given the lifetime of the sub-giant 
phase  derived from  theoretical models, this solution  is quite unlikely \citep{Andersen90}.  Furthermore, as shown 
by \citet{Boden05}, the evolutionary state of the fitted models  is highly dependent on the model details (and in 
particular on the amount of core overshooting) and needs further investigation. The masses of 12 Boo components 
(1.416 and 1.372 \msol) are in fact in the transition region, where the thickness of the ``overshooting''  layer seems to 
depend on the stellar mass, metallicity and evolutionary stage \citep{Vandenberg06}.

Even if the masses are very precisely known, and both the components have the same age and initial chemical composition,
there is a large number of possible parameter sets able to fit the ``classical'' observables.
In this paper we present different scenarios based on different theoretical prescriptions for extra-mixing processes,
which provide stellar parameters compatibles with observational constraints, but with different evolutionary state 
and internal structure.
The classical observables  do not allow to distinguish between different solutions, and hence
additional and independent observational constraints are needed.
In this paper we propose solar-like oscillations as additional observational constraints. Fortunately, the components of 12~Boo are located in a region where solar-like oscillations are expected to be excited (see e.g. \citealt{Kjeldsen95}, \citealt{Houdek99} and  \citealt{Samadi05}).

 Solar-like oscillations have been observed in several single stars 
\citep[e.g. $\eta$ Bo\"otis, and $\mu$ Arae,][ respectively]{Kjeldsen95,Bouchy05} 
and binary  systems  \citep[Procyon,][]{Martic04}  as well as in both  components 
  of the visual binary system $\alpha$ Centauri \citep{Bouchy02, Carrier03, Bedding04, Kjeldsen05}.  In the latter, 
 as shown by \citet{Miglio05}, the combination of asteroseismic and precise ``classical'' constraints (masses, radii, 
 luminosities \ldots) significantly improves the determination of the system fundamental parameters.  While the 
 study of $\alpha$ Centauri provides a test of our knowledge of stellar structure in conditions that are slightly
  different from the Sun,  a detailed investigation of 12 Bo\"otis will be a relevant step in understanding  
 the structure of the central regions for this particular area in the  Hertzprung-Russell  diagram (HRD)
  where the  ``transition'' 
from radiative to convective core models takes place.

In Sect.~\ref{sec:constraints} we summarize the observational knowledge about 12 Boo, and we describe the
constraints will be used in the calibration process. The latter is described in  Sec.~\ref{sec:modelling}.
Among the free parameters, the most important is the extension of the overshooting region.  
We will also analyse possible  solutions  by taking
into account different physical processes able to change the thickness and features of the extra mixed region.
In Sec.~\ref{sec:seismology} we study how the differences in the stellar structure for the models
fitted in different scenarios are reflected on the oscillation frequencies, and we suggest which kind of
observational data could give us enough information to constraint the evolutionary state of 12 Boo,
and, possibly, also the properties of the transport processes (if any) in the stellar central region.
Finally, in Sec.~\ref{sec:conclusion} we present our conclusions.

\section{12 Bo\"otis: observational constraints} 
\label{sec:constraints}
With the rapid development of optical interferometry in the last years, closer and
closer binary systems have become resolvable.  Thanks to interferometric
orbits an increasing sample of double-lined spectroscopic 
binaries (SB2), including non-eclipsing systems, can  be used for accurate   
stellar parameter determination. Interferometry essentially yields the orbital inclination 
that, combined with accurate radial velocity measurements, can provide
mass determination at a level better than 1\%, i.e. suitable for strict stellar 
modelling tests \citep[see e.g.][]{Andersen91}.

A good example of an ideal target for this technique is  12~Boo,  a bright (V=4.83) non-eclipsing SB2 with an orbital 
period of about ten days ($P=9.6$) and a (composite) spectral type  F8 IV -F9 IVw \citet{Barry70} (where  ``w" 
stays for weak metallic lines).
\citet{Boden05} obtained long baseline interferometry of 12 Boo with the Palomar
Testbed Interferometer (PTI)  and the Navy Prototype Optical Interferometer (NPOI).
 They  secured, as well, new high resolution (echelle) spectra, as the previous spectroscopic observations 
 \citep{Abt76,Demedeiros99}, were the major limit to the precision of derived parameters.
 The interferometric visibility data and the radial velocity curves were simultaneously
solved to derive the orbital parameters. The  components could not be 
resolved (the apparent orbital semiaxis is only 3.45 mas), so that the individual radii 
had to be derived by indirect means, namely by the Infrared Flux Method by Blackwell
and collaborators \citep{Blackwell90,Blackwell94}. The information about the total
and the individual bolometric fluxes   (needed  for radius determination) was derived 
by a fit of multi-wavelength archival data and from the in-band intensity ratio of
the interferometric measurements.  Furthermore, the analysis of the observed spectra 
provided an estimation of the component effective temperatures and of the luminosity
ratio. 
A further improvement in the determination of the spectroscopic orbit 
by \citet{Tomkin06}, combined with the same interferometric data, confirms the values of the 
masses derived by \citet{Boden05} and reduces the associated error bars.
The relevant values and their uncertainties are reported, for the reader's sake,
in Table \ref{orbit}.

Other relevant information on the system are its near-solar chemical composition,
with a mean value of [Fe/H]$=0.01 \pm 0.08$, derived from Str\"{o}mgren photometry and 
detailed spectroscopic analysis \citep{Duncan81, Balachandran90, Lebre99, Nordstrom04} 
and the component rotational velocities of 14.0 and 12.0~km~s$^{-1}$ \citep{Boden05}. 
The latter are slightly faster than the expected corotation rate of, respectively, 12.4 
and 9.3~km~s$^{-1}$. 

The resulting global picture of 12 Boo is that of a binary with very similar 
components (masses,effective temperatures, rotation) except an 
unexpected magnitude difference: both spectroscopic  and interferometric measurements 
indicate  a difference of $\sim$ 0.48-0.58~mag, depending on the band. That is
quite hard to explain in a system with coeval and non-interacting components 
(both of them are well inside the critical Roche lobe).

The high precision of the radial velocity measurements \citep[the typical error is  
0.5~km~s$^{-1}$ in][]{Boden05,Tomkin06} and the  short orbital period 
rules out the possibility of making the masses unequal by ''hiding" a small fraction 
of a component mass in a third unseen object. Unless rather improbable configuration 
of the third object orbit are assumed (i.e. orbit almost normal to the line of sight), even 
an object of a few hundredth solar masses orbiting one component would easily be detected
from the reflex motion of the more massive companion. Besides, such a planetary orbit
can be excluded on the ground of stability consideration, see e.g  \citet{Holman99} and 
references therein.

\begin{table}
\begin{minipage}[t]{\columnwidth}
\caption{Orbital parameters and magnitude differences for 12 Boo}
\label{orbit}
\centering
\renewcommand{\footnoterule}{}  
\begin{tabular}{lcc}
\hline
Parameter  & value & reference \footnote{References: 1) \citet{Tomkin06},2) \citet{Boden05}} \\
\hline\hline

$P$ (d)									& 9.6045529 $\pm 4.8\, 10^{-6}$	& 1	\\
$e$										& 0.19268   $\pm$ 0.00042		& 1	\\
$i$										& 107\fdg 99 $\pm$ 0.077		& 2	\\
$a$ (mas)								& 3.451   $\pm$ 0.018			& 2	\\ 
$\Delta K_{\mathrm{CIT}}$ (mag) 		& 0.589   $\pm$ 0.005		 	& 2	\\
$\Delta H_{\mathrm{CIT}}$ (mag) 		& 0.560   $\pm$ 0.020			& 2	\\
$\Delta V$ (mag)						& 0.560   $\pm$ 0.020			& 2	\\
$K_{\mathrm{A}}$ (Km~s$^{\mathrm{-1}}$) 	& 67.286  $\pm$ 0.037 		& 1	\\ 
$K_{\mathrm{B}}$ (Km~s$^{\mathrm{-1}}$) 	& 69.30   $\pm$ 0.050			& 1	\\ 
\hline
\end{tabular}
\end{minipage}
\end{table}

The observational constraints we assume in our modelling are taken from Table~5 of \citet{Boden05} that, for the  
sake  of convenience, we repeat here in Table~\ref{tab:obs}. As the above mentioned spectroscopic and photometric    
studies of 12 Bo\"otis  indicate a metallicity within 0.1 dex of the solar value,  given this uncertainty and a  
further uncertainty on the value of the solar metallicity, we adopted the solar
 $Z/X$ of \citet{Grevesse93} considering a conservative error bar, as reported in Table \ref{tab:obs}.

\begin{table}
\centering
\caption{\small Observational constraints as adopted in the modelling.}
\label{tab:obs}
\begin{tabular}{lcc}
  \hline
  & A & B \\
\hline
\hline
M/\msol & 1.4160$\pm$0.0049& 1.3740$\pm$0.0045\\
\teff~(K)& 6130$\pm$100 & 6230$\pm$150 \\
L/\lsol & 7.76$\pm$0.35 & 4.69$\pm$0.74\\
Z/X & 0.0245$\pm$0.01 & 0.0245$\pm$0.01 \\
 \hline
 \end{tabular}
\end{table}

\section{Modelling 12 Bo\"otis} 
\label{sec:modelling}
In the following we will use the term ``calibration'' to denote the process of determining the
stellar model parameters, such as initial composition and age (both the same for A and B components), 
  satisfying - within 1-$\sigma$ -
the  observational constraints: $M_{\rm A}$, $M_{\rm B}$, $Z/X$, $L_{\rm A}$, $L_{\rm B}$, $T_{\rm eff,A}$ 
and $T_{\rm eff,B}$. 

As both system components  are massive enough to develop a convective core during the main-sequence and  
no a-priori assumption on their evolutionary state can be made, the dependence of \teff\ and luminosity 
 on the stellar parameters (e.g.  age and overshooting) is highly non-linear. 
As \citet{Boden05} already pointed out, the result of  system fitting  strongly depends  on the  model details,  
and, in particular, on the amount of convective overshooting in the core.

The aim of this paper is to study the possible solutions (not just to find a single solution)  
and to verify 
the ability of discriminating among different scenarios.  
We apply, therefore, the minimization technique adopted and  described in \citet{Miglio05} with the purpose of  
determining the chemical composition and age for each value of the overshooting parameter (assumed to vary from   
0.0 to 0.50 in steps of 0.01).
The parameters left free 
are: the initial chemical composition (described by the hydrogen and heavy-elements mass fraction, X and Z  
respectively), 
the age and the component masses. 
These  are adjusted in order to fulfill the observational constraints  in Table~\ref{tab:obs}. 
A standard  $\chi^2$ is used as  goodness-of-fit estimate, so that each observable is taken into account 
with its uncertainty. 
We did not allow  $\alpha_{\rm MLT}$  to be adjusted and fixed its value to  $\alpha_{\rm MLT}=1.84$,
i.e. the value provided by the Solar calibration.
This choice is justified by the fact that  both stars  are quite  close 
in the HRD and  that different values of  $\alpha_{\rm MLT}$  are not expected. 
Besides, a variation of $\alpha_{\rm MLT}$ implies a change of \teff\ 
that can be easily balanced by a change of the chemical composition. 

The search for a solution fulfilling the observational constraints on the
HR diagram is usually biased by an additional criterion, though this is
not included in the  $\chi^2$: 
 a preference towards long lasting evolutionary phases, as the probability of finding a star in a given 
evolutionary phase depends on how ``rapid'' that phase is. 
Evidently,  it is less likely to observe a star in 
its second gravitational contraction than on the main sequence.  This is
however a statistical argument, that can be verified on 
large samples of stars (e.g. in clusters), but cannot be applied  
to a  single case. 
In addition to that, the duration of each evolutionary stage depends on whether overshooting is considered in the models or not.
As pointed out e.g. by \citet{Maeder75} and \citet{Noels04}, the duration of the 
thick-shell-hydrogen burning phase   depends on the amount of overshooting as the latter increases the 
mass of the isothermal helium core built during the core hydrogen-burning phase. 
As an example, a 1.4\msol\, star without overshooting spends in that phase
 $\sim 20\%$ of the main-sequence lifetime \citep[see][ for instance]{Iben91}. An overshooting parameter 
$\alpha_{\rm OV} \simeq 0.1$ implies a  decrease of the duration of this evolutionary phase 
down to $4 \%$ of the main-sequence lifetime, i.e. lasting as long as the the overall contraction phase.
In our modelling we  do not make, therefore, any a-priori assumption on the particular evolutionary phase 
of the solution.  

\subsection*{Stellar Models}
The stellar model sequences are computed with the code CLES  (Code Li\'egeois d'Evolution  Stellaire).  The  opacity 
tables are those  of OPAL96 \citep{Iglesias96} complemented at $T < 10000$~K with \citet{Alexander94}  
opacities. 
The  metal  mixture  used in the opacity tables, is the solar one according  to   
\citet{Grevesse93}.
The nuclear energy generation routines are based on the cross sections by \citet{Caughlan88} and
updated with the recent measurement of the $^{14}N(p,\gamma)^{15}O$ reaction rate \citep{Formicola04}.
The weak screening factors come from \citet{Salpeter54}, and 
the equation of state used in the computations is  OPAL01 \citep{Rogers02}.
Convection transport is treated with the classical
mixing length theory by \cite{Bohm58} in the formulation by \cite{cox}; and  atmospheric boundary conditions given by 
\cite{kurucz98} are applied at $T=T_{\rm eff}$.

Chemical mixing in the formal convective region and in the overshooting layer is treated as instantaneous. Besides,
the thickness of the overshooting layer, $\Lambda_{\rm OV}$, is expressed in terms of a parameter $\alpha_{\rm  
OV}$, 
so that  $\Lambda_{\rm OV}=\alpha_{\rm OV}\times\min(r_{\rm cv}, H_{\rm p}(r_{\rm cc}))$, where $r_{\rm cc}$ is  
the radius of the convective core.
The temperature stratification in this overshooting layer could be radiative or adiabatic (e.g. Zahn 1991).
 At any rate the effect of this thin layer on the evolutionary track is  very small, in our models a radiative
 stratification has been assumed.

We have computed as well models that, in contrast with the assumed instantaneous chemical mixing, include a
diffusive treatment of mixing in the formal convective region and in the extra-mixed layer.
These computations were done with the ATON3.0 code \citep{Dantona05} for the same physics: MLT, OPAL1996 
and Alexander \& Ferguson opacities, as well as OPAL equation of state. 

We have also studied (see Sect.~\ref{sec:diffusion}) the effect of other chemical mixing processes, such as diffusion and 
gravitational settling of helium and  heavy elements.
Microscopic diffusion has been implemented in CLES code following the formulation by \citet{Thoul94}.

Finally, the effect of an additional turbulent mixing, such as the one expected to be induced by rotation, is 
addressed in Sec. \ref{sec:diffov}.
\begin{table}
\centering
\caption{\small Parameters of the models.}
\label{tab:results}
\begin{tabular}{lcccc}
\hline
&  $\alpha_{\rm OV}$ & $Y_0$ & $Z_0$ & Age (Gyr) \\
\hline
a  & 0.06        & 0.268 & 0.0195 & 3.08\\
b  & 0.15        & 0.267 & 0.0198 & 3.31\\
c  & 0.23        & 0.267 & 0.0200 & 3.58\\
d  & 0.37, 0.15  & 0.281 & 0.0186 & 3.24\\
\hline
 \end{tabular}
\end{table}


\subsection{Overshooting} 
\label{sec:amountov}
\begin{figure*}
\begin{center}
\resizebox{0.6\hsize}{!}{\includegraphics[]{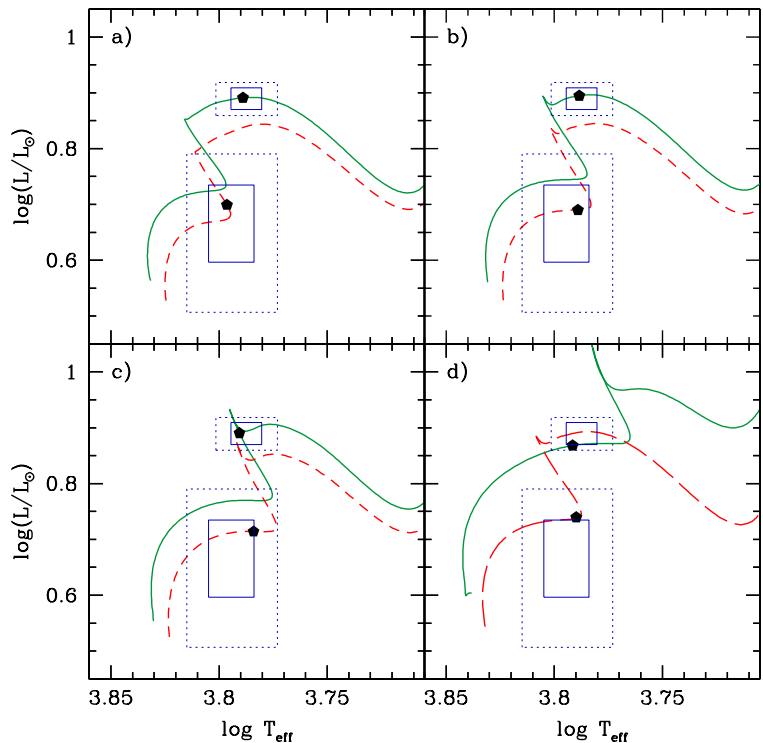}}
\caption{\small HR diagram showing solutions found with different values of the overshooting parameter: {\it a)} 
$\alpha_{\rm OV} = 0.06$, {\it b)} $\alpha_{\rm OV} = 0.15$, {\it c)} $\alpha_{\rm OV} = 0.23$, and {\it d)}  
$\alpha_{\rm OV,A} = 0.37$ and $\alpha_{\rm OV,B} = 0.15$. Model parameters are described in Table  
\ref{tab:results}. Error boxes corresponding to 1- and 2-$\sigma$  error bars in $L$ and \teff\ are represented by continuous and dotted lines.}
\label{fig:solution1}
\end{center}
\end{figure*}

We performed several calibrations for different values of the overshooting parameter and we found  solutions  with  
$\alpha_{\rm OV}\geq 0.06$ (see. Table~\ref{tab:results}).
By changing  $\alpha_{\rm OV}$ the evolutionary state of the system changes. In detail: for the lowest  
values of overshooting the situation is that  represented in Fig.~\ref{fig:solution1}.a: the primary component 
is close to the  maximum luminosity in the shell--hydrogen burning phase, whereas the secondary, with a 
$\rm X_c$=0.03, is at the end of its central  hydrogen-burning phase.
As  $\alpha_{\rm OV}$ increases, the evolutionary state of  component A approaches the TAMS. With $\alpha_{\rm  
OV} \simeq 0.15$ (Fig.~\ref{fig:solution1}.b) the primary is burning hydrogen in a thick shell   
while  component B is well on the main sequence. A  further increase of the overshooting parameter 
 $\alpha_{\rm OV} \geq 0.2$ places the primary in the rapid overall contraction phase (Fig.~\ref{fig:solution1}.c).
Still larger values of $\alpha_{\rm OV}$ make $L_{\rm B}$ increase out of 1-$\sigma$ and therefore worsen the fit.

 The observed luminosity ratio,
$L_{\rm A}/L_{\rm B} \sim 1.65$ (from a minimum of 1.35 to a maximum of 2.)
is the main obstacle for a solution with both components on the Main-Sequence. In fact, given that
both components have the same age and  initial chemical composition, if the same kind of mixing processes are  
considered in the center, 
the luminosity ratio in the main-sequence is determined  by  
the  mass ratio ($q\sim0.97$), and does not significantly depend on the choice of the other common 
model parameters  ($X,Z,\alpha_{\rm OV}$).

Regardless of the set ($X,Z,\alpha_{\rm OV}$) chosen, all the models  computed with 
 $\alpha_{\rm OV,A}=\alpha_{\rm OV,B}$ provided $(L_{\rm A}/L_{\rm B})_{\rm MS} \simeq 1.15$. 
A higher value of the luminosity ratio can be reached if  $\alpha_{\rm OV,A} > \alpha_{\rm OV,B}$.
Therefore, the only way we found to place both components in their MS phase is to assume 
a different efficiency  of  mixing processes in each component. 
In Fig.~\ref{fig:solution1}.d  we show a calibration solution  obtained by assuming  
$\alpha_{\rm OV, A} \simeq 0.37$
 and $\alpha_{\rm OV, B} \simeq 0.15$. In this case both  12 Boo A and B are well in  the MS.

The values of stellar parameters collected in Table~\ref{tab:results} are those 
used in the figures and comparisons in this paper. It has to be noticed, however, that the uncertainty 
on $Z/X$ reflects on an undertainty of 0.02 in  
the initial  helium mass fraction ($Y_0$)  and of  0.2~Gyr in the age of the models. 
Besides, the calibrations with $\alpha_{\rm OV,A}=\alpha_{\rm OV,B}$  and  similar initial chemical 
composition provide an age  difference  of the order of 15\%, and the 
 calibration with $\alpha_{\rm OV,A} > \alpha_{\rm OV,B}$
implies an age in the range bracketed by the $\alpha_{\rm OV,A}=\alpha_{\rm OV,B}$ models.
Unfortunately, the uncertainty in Z/X is too large to constrain the models.

All the different scenarios  presented here cannot be  clearly discriminated by means of the available 
observational constraints,
nevertheless, the internal structure of these models are significantly different, and these  
differences  shall show up in  their asteroseismic properties (see. Sec.~\ref{sec:seismology}).

\subsubsection*{Diffusive overshooting}
The classical treatment of overshooting is an extension
of the instantaneously mixed convective core  by a distance $\Lambda_{\rm OV}$.
In the diffusive approach of convection, chemical mixing is described by the diffusion equation with diffusion  
coefficient $D=1/3\,v_{\rm conv}\Lambda$, where $v_{\rm conv}$ is the average turbulent velocity and $\Lambda$
 the convective  scale length. In
a MLT treatment $v_{\rm conv}\propto (\nabla-\nabla_{\rm ad})$ and $\Lambda\propto H_{\rm p}$, which imply
 $v_{\rm conv}=0$ at the 
formal boundary of the convective region ($\nabla=\nabla_{\rm ad}$). In order to 
define a continuous diffusion coefficient, also in the
overshooting region, ATON code extrapolates the $\log v_{\rm conv}\, \,vs.\,\log P$ relation,
to get the turbulent velocity at the boundary, starting from a point inside the convective region 
where the pressure is 5\% larger than at the boundary. 
Furthermore, following \citet{Xiong85},  turbulent velocity is assumed to exponentially vanish outside the 
convective region, and the thickness of the mixed region is determined by the parameter $\beta$
which describes the exponential vanishing of turbulent velocity. As  shown by \citet{Ventura98}, a 
$\beta$ value ten times smaller than the $\alpha_{\rm OV}$ in classical instantaneous mixing produces
the same effect on  the HR diagram.

With this formulation of overshooting we find the same set of solutions as with instantaneous mixing, this time
 for a parameter $\beta$ in the range 0.004-0.035. This is of course reassuring as a completely independent stellar
 evolution code, and a different description of overshooting have been used.
The chemical composition profiles at the border of the convective core are, however, much
smoother with the diffusive overshooting than with the classical treatment.

\subsection{Effects of diffusion}
\label{sec:diffusion}
 \citet{Richard01}, \citet{Michaud04} and \citet{Richard05} showed that microscopic diffusion can affect the size  
and evolution of a convective core. In those papers, thermal diffusion, gravitational settling and radiative 
accelerations were included in the models  to derive the evolution  of chemical distribution.
CLES code does not take radiative accelerations into account. 
Nevertheless, there is no observational evidence in 12~Boo of surface chemical peculiarities that could be  
produced by   gravitational-settling and radiative-acceleration,
and the latter is  not expected  to play a mayor role in the central region \citep[see e.g.][]{Richard01}.

We do not expect a large effect of radiative acceleration in the 12~Boo mass domain,  
since the thickness of their convective envelopes is quite large. 
Even if at the beginning of the MS  component A has a shallow convective envelope, and therefore the 
role of radiative accelerations could be non-negligible, as the star evolves its convective region becomes deeper. 
As  a consequence, the
effect of radiative acceleration is reduced, and the chemical composition of the convective region re-homogenized.
We  decided, therefore, to consider and compare models accounting for diffusion of H and He, and of H, He and Z.
  
\citet{Michaud04} showed that  microscopic diffusion can induce an increase of the convective core for a narrow  
range  of masses, from 1.1 to 1.5 \msun, and that the effect decreases rapidly with increasing  stellar mass.
 We  also note that in this mass domain, the convective core mass increases during the MS evolution 
instead of decreasing, as it occurs for larger masses. As a consequence, a sharp gradient of chemical composition 
appears at the convective core border, making the diffusion process much more efficient in that region.

\begin{figure}
\begin{center}
\resizebox{0.8\hsize}{!}{\includegraphics[]{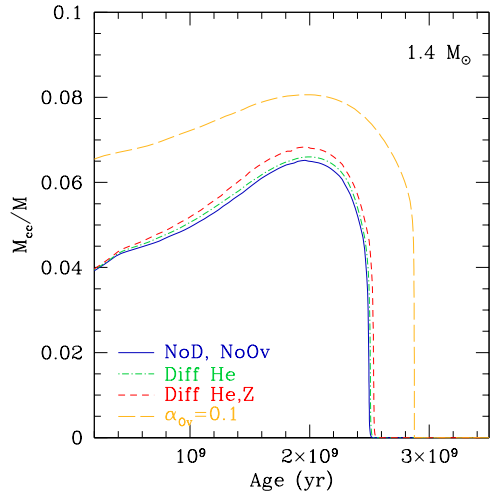}}
\caption{\small Fractional mass of the convective core as a function of the age 
in no diffusion models: without overshooting (solid line) and with $\alpha_{\rm ov}=0.1$ (long-dashed line);
and in  no overshooting models including microscopic diffusion of He and H (dashed-dotted line) and of
He,H and Z (dashed line).}
\label{fig:agemcc}
\end{center}
\end{figure}

\begin{figure}
\begin{center}
\resizebox{0.8\hsize}{!}{\includegraphics[angle=0]{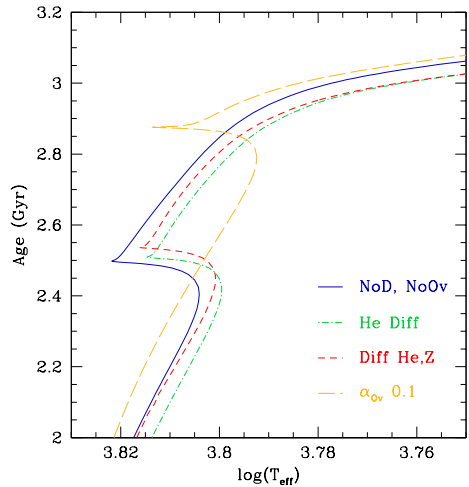}}
\caption{\small  Evolutionary tracks of the models described in Fig. \ref{fig:agemcc}
 in a $\log{T_{\rm eff}}$-Age  diagram. }
\label{fig:duratad}
\end{center}
\end{figure}

As  shown in Fig.~\ref{fig:agemcc}, for a 1.4~\msun\ star, the increase of the core mass fraction when diffusion  
of H, He and Z is considered is of only $\simeq 5\%$, and the age of the turnoff is only slightly increased
 (Fig.~\ref{fig:duratad}).  This is in agreement with the findings of \citet{Michaud04} and \citet{Richard05} who  
consistently  included radiative accelerations in their computations while considering  $Z$-diffusion.
 An obvious consequence is that, in presence of diffusion, the amount of overshooting 
needed  to find 12~Boo~A and B in a given evolutionary state is reduced with respect to the values
obtained in the previous section. In fact, the minimum amount of overshooting required is slightly smaller, 
i.e. $\alpha_{\rm OV} \simeq  0.04$ instead of 0.06.  The evolutionary state of the calibrated system changes, 
with increasing  $\alpha_{\rm OV}$, in a way similar to  that described Sec.~ \ref{sec:amountov}.

On the other hand, the increase of the opacity in the envelope due to the settling 
of He and Z provides  evolutionary tracks  cooler than 
those without diffusion (this effect is even more important if only H and He diffusion is
considered, see Fig. \ref{fig:duratad}). At the end of the main sequence the surface $Z/X$ is $\sim 10 \%$  
(diffusion of H and He) or $\sim 25 \%$ 
(diffusion of H, He and Z) smaller than the initial value, nevertherless, 
given the uncertainties in the observed $Z/X$, this   implies only a slight change 
in the  $Z$ and $X$ initial values in the calibrated models.

In addition to a slight enlargement of the convective core, 
diffusion leads to  a smoother chemical composition profile at the edge of the convective core (see Fig. \ref{fig:core-dr}). 
Whether this property affects the oscillation modes of the components will be addressed in see Sec.  
\ref{sec:seismology}.

\subsection{Other sources of extra-mixing in the core}
\label{sec:diffov}

The shear instability induced  by stellar rotation is considered as one of
the first causes of chemical mixing in the radiative interior of the stars.
Different approaches have been proposed to treat the effects of rotation on transport
of angular momentum and chemicals \citep[see][ for a review]{Maeder00, Mathis04}.
Since our stellar evolution code does not include a consistent treatment of rotational effects, 
we simply include the chemical turbulent mixing by adding a turbulent diffusion coefficient
($D_{\rm T}$) in the diffusion equation. In our parametric approach $D_{\rm T}$ is assumed to
 be constant inside the star and independent of  age. These simple prescriptions are inspired by
 the results obtained by \cite{Mathis04} from a consistent study of 
rotational effects in a 1.5~\msun\ star (see their Fig. 2).

We have then calibrated our models with different values of $D_{\rm T}$. We find, similarly 
to the case of including overshooting (see Sec. \ref{sec:amountov}), that we are able to fit the system at
 different evolutionary stages depending on the value of $D_{\rm T}$; 
the values needed to fit the system are in the range $30-170$~cm$^2$ s$^{-1}$. We note that these values are
of the same order of the total diffusion coefficient 
resulting from the calculations for a 1.5 \msun\ model as presented in \cite{Mathis04}.
As already seen for overshooting, 
in order to fit the system with 12 Boo~A and B in their main-sequence, a different efficiency 
of mixing must be assumed in each component: $D_{\rm T,A}=330$~cm$^2$ s$^{-1}$, and $D_{\rm T,B}=120$~cm$^2$ s$^{-1}$.

 The simplified parametric treatment of rotationally induced mixing used in this work has the aim of 
showing that if an extra-mixing process, different from overshooting, is acting near the core it will 
produce a different chemical composition profile in the central regions of the star. 
Whether these different profiles are reflected or not  on the solar-like oscillations features will be  analysed
in Sec.\ref{sec:profile}.

\section{Asteroseismic properties}
\label{sec:seismology}

Stellar models predict the presence of a convective envelope for the range of masses of 12 Boo~A and B
and, therefore,   solar-like oscillations are expected to be excited in both components of the system.
By means of the scaling relations of \cite{Kjeldsen95}, \cite{Kjeldsen01} and the theoretical predictions 
of \citet{Houdek99}, and  using the precisely determined values of the mass and the luminosity, we  estimate that
 the modes  with the largest amplitudes should be around $\nu_{\rm max}=700$~$\mu$Hz  and  
$\nu_{\rm max}=$1200~$\mu$Hz in 12 Boo A and B, 
respectively. The peak amplitude in the power-spectrum is expected to be 3-4.5 times larger than the solar-one for
 the A component, and 2.5-3.5 times larger then the solar one for the B component. 

The components of 12 Bo\"otis are massive enough to develop a convective core during  main sequence. The combined  
action of nuclear burning and 
convective mixing is responsible for the development of a steep chemical composition gradient at the boundary of  
the convective core.
As a star leaves the main-sequence, the increasing central condensation, together with the steep chemical  
composition gradient in the central regions, 
leads to a large increase of the buoyancy frequency ($N$) and, therefore, of the frequencies of gravity modes. The  
latter interact with pressure modes and 
affect the properties of non-radial solar-like oscillations by the so-called {\it avoided crossing} phenomenon  
\citep[see e.g. ][]{Osaki75,Aizenman77}. The frequency of the modes undergoing an {\it avoided crossing} 
(also called mixed-modes) are therefore sensitive probes of the core structure of the star.

\begin{figure}
\begin{center}
\resizebox{0.9\hsize}{!}{\includegraphics[angle=0]{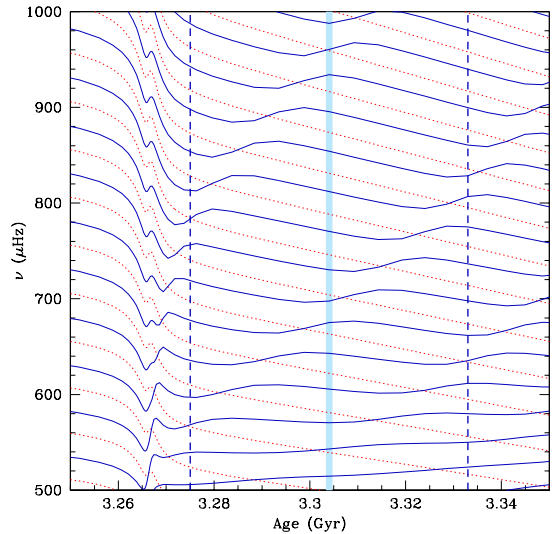}}
\caption{\small {Evolution in time of the frequencies of model A$b$. Dotted and solid lines represent radial and  
$\ell=1$ modes with radial order $n$ from 11 to 23.
The frequencies of model A$b$ are highlighted by a thick vertical line. Vertical dashed  
lines represent $\pm1\,\sigma$ interval in the observed $T_{\rm eff}$.}}
\label{fig:evolfr}
\end{center}
\end{figure}

As a first example let us consider the evolution of the frequencies of the model of component A 
corresponding the scenario $b$ of Table~\ref{tab:results} (hereafter A$b$). In 
Fig.~\ref{fig:evolfr} we plot the evolution of frequencies for $\ell=0$ (dotted lines) and $\ell=1$ (solid lines)
modes, and   show how $g$-modes, whose frequency strongly increases after the overall contraction
 ($\sim 3.27$~Gyr), interact   with pressure modes of same degree. 
This interaction results in non-radial  modes with a mixed $p-$ and $g-$ character  and frequencies 
that significantly deviate  from a regular spacing between consecutive overtones.

\begin{figure}
\begin{center}
\resizebox{0.9\hsize}{!}{\includegraphics[angle=0]{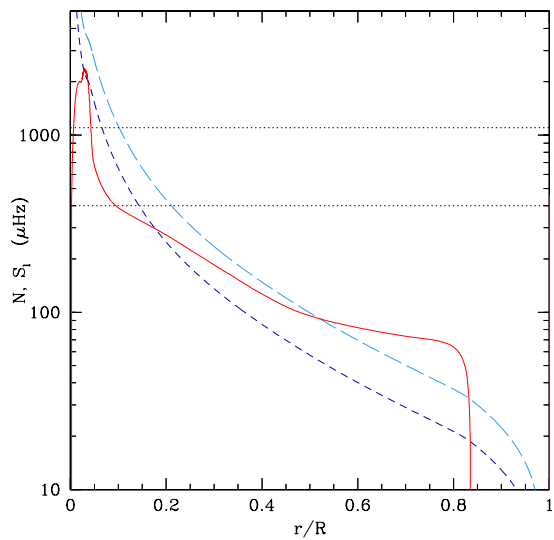}}
\caption{\small{Propagation diagram for model A$b$ (see Table \ref{tab:results}, Fig.~\ref{fig:solution1}). The solid  
line represents the \BV 
frequency $N$, the short- and long-dashed line the Lamb frequency $S_\ell$ for, respectively $\ell=1$ and $\ell=2$. The domain  
of solar-like oscillations 
falls within the horizontal dotted lines.}}
\label{fig:na}
\end{center}
\end{figure}

Since the thickness of the mode evanescent region 
(where $N < \omega < S_\ell$, with $S_\ell$ being  the so-called Lamb frequency)
increases with the degree $\ell$ (see propagation diagram in Fig.~\ref{fig:na}), the interaction
 between p and g modes decreases.  As a consequence, in modes of 
degree $\ell>1$, the behaviour of gravity and pressure mode is generally better  separated 
\citep[see e.g.][]{Jcd95b, Morel01}:  g-modes tend to remain concentrated in the core 
and would hardly be detectable due to their significantly larger inertia.
This is in fact the case for the our models A{\it a} and A{\it b} where only $\ell=1$ modes
present a mixed character. Nevertheless, for the less evolved models (A{\it c} for instance)
   also  $\ell=2$ mixed modes appear in the solar-like frequency domain.
In the following discussion we will
address mainly the properties of mixed modes.

Given the MS status of 12 Boo~B, we will first discuss the seismic properties of
component A,  and only in Sec. \ref{sec:seismoB} consider the additional information that the 
observation of solar-like oscillations of component B would add to the modelling.

\subsection{Probing the evolutionary status}
As already investigated in the case of $\eta$ Bo\"otis \citep[see e.g.][]{Dimauro04}, the frequencies of modes  
undergoing an {\it avoided-crossing}
are direct probes of the evolutionary status of intermediate mass stars.

\begin{figure}
\begin{center}
\resizebox{\hsize}{!}{\includegraphics[angle=0]{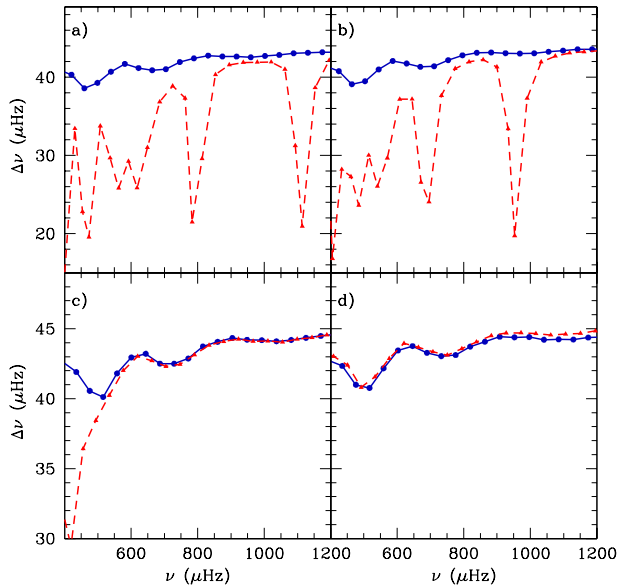}}
\caption{\small Large frequency separation as a function of the frequency for the models described in 
Fig.~\ref{fig:solution1}. Solid and dashed 
lines connect, respectively, $\Delta\nu$ for radial and $\ell=1$ modes. 
}
\label{fig:ls}
\end{center}
\end{figure}

In Fig.~\ref{fig:ls} we compare the large frequency separation\footnote{Computed as the difference between modes  
of same degree and consecutive 
frequencies.} $\Delta\nu$ for radial and $\ell=1$ modes corresponding to  the models of component A 
 presented in Fig.~\ref{fig:solution1}.
Though radial modes do not give information on the evolutionary status, $\ell=1$ modes allow a clear  
discrimination among the scenarios.
Modes of mixed p and g character are not present in the model on  
the main sequence (Fig.~\ref{fig:ls}.d) and $\Delta\nu$ is almost constant,
indicating therefore a regular spacing between modes of consecutive radial order (both radial and $\ell=1$).
On the other hand, as more evolved models (A$a$, A$b$ and A$c$) are considered, the
 {\it avoided crossings} effects become more evident. Therefore, for the model in the second overall contraction 
 (A$c$, Fig.~\ref{fig:ls}.c) the interaction between  p and g modes changes by up to 10~$\mu$Hz 
the modes with frequencies lower than 600~$\mu$Hz (corresponding order-mode, $n$, lower than 12).
 That can  also be  seen in Fig.~\ref{fig:evolfr}  for models at  3.265~Gyr and the lowest $n$ modes.
For more evolved models (A$a$ and A$b$) the avoided crossing phenomenon clearly breaks the regular frequency
spacing in  the domain of expected solar-like oscillation  (Figs.~\ref{fig:ls}.a and ~\ref{fig:ls}.b).

\begin{figure}
\begin{center}
\resizebox{\hsize}{!}{\includegraphics[]{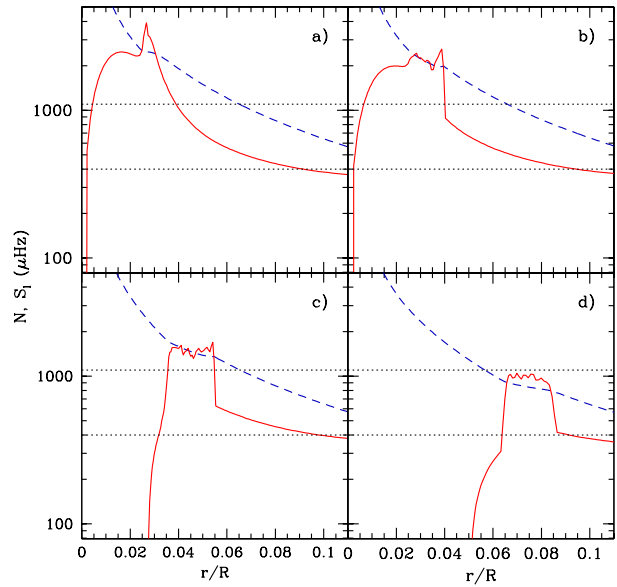}}
\caption{\small Propagation diagram in the core of the models for which large frequency separation has been
plotted  in Fig.~\ref{fig:ls}.
Solid line represents the \BV frequency, and  the  dashed line the Lamb frequency for $\ell=1$.}
\label{fig:nall}
\end{center}
\end{figure}

The radical variation of the properties of $\ell=1$ modes derives from a considerable change of the \BV frequency  
near the center of the star (see Fig.~\ref{fig:nall}) as hydrogen is exhausted throughout the convective core and the star undergoes an
overall contraction. As models of component A at different evolutionary stages are obtained for different amounts of  core overshooting, 
 the detection of mixed modes  and their frequencies would allow to constraint the
value of the overshooting parameter or the  extension of the mixed central region.
While model A$a$ (computed with $\alpha_{\rm OV}=0.06$)  presents  {\it avoided crossings} at $\nu=780\,\mu$Hz ($n=15$)
and  $\nu=1100\;\mu$Hz ($n=$24)  model A$b$ ($\alpha_{\rm OV}=0.15$) does it at  $\nu=700\;\mu$Hz ($n$=13)
and  $\nu=950\;\mu$Hz ($n$=20). These differences between the frequencies of the modes presenting mixed p-g character ($\sim 100\;\mu$Hz),
as well as the differences of the frequency spacing between two consecutive avoided crossings ($\sim 20\,\mu$Hz), 
are large enough to be  detected with the current observational techniques.
Furthermore, 12 Boo~A has a rotational velocity ~14~km s$^{-1}$ \citep{Boden05} that, assuming an inclination of the rotational axis equal to  the orbital one (108$\degr$), implies a rotational splitting between $(\ell=1,m=-1)$ and   $(\ell=1,m=1)$ modes of only  $2.5\;\mu$Hz.

Distinguishing between scenarios  A$c$ ($\alpha_{\rm OV}=0.23$) and A$d$ ($\alpha_{\rm OV}=0.37$ 
for 12 Boo~A and $\alpha_{\rm OV}=0.15$ for 12 Boo~B) on the base of $\ell=1$ mixed modes 
might be less straightforward since the effect of avoided crossing occurs at  lowest frequencies, where the amplitudes of the
modes are predicted to be small. Therefore, no
detection of avoided crossing could imply, either that 12 Boo~A is well  in MS, and  has  therefore a large 
central mixed region, or that  it is close to the TAMS.  
Luckily, for models in the second overall contraction (SOC), the changes in  stellar structure are
such that $\ell=2$ mixed modes appear in the predicted solar-like spectrum, while for MS models (A{\it d})
they would have very low frequencies.

\subsubsection{Seismology of component B}
\label{sec:seismoB}

\begin{figure}
\begin{center}
\resizebox{.8\hsize}{!}{\includegraphics[]{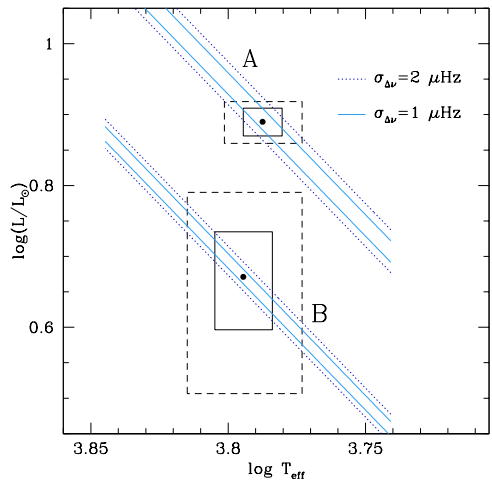}}
\caption{\small  Position in the HR diagram of 12 Boo A and B with 1
and 2-$\sigma$ error boxes in $\log{L}$ and $\log{T_{\rm eff}}$. Also
shown are  lines of constant radius deduced from the (known) masses of
the components and the large frequency separation. The large
separation is assumed to be known with an uncertainty of 1 (continuous
lines) and 2 $\mu$Hz (dotted lines).}
\label{fig:radiusB}
\end{center}
\end{figure}

Though in all  calibrations we considered component B as a
main-sequence object and, therefore, avoided crossings are not expected,
the detection of solar-like oscillations in the secondary would,
nonetheless, add constraints to the modelling of the binary system. A first
additional constraint comes from the average value of the large
frequency separation ($\langle \Delta\nu\rangle$) that, for
main-sequence stars, is well known to represent an estimate of the
mean density and, hence, of the stellar radius, if the mass is known.
The available estimate of  component radii in 12~Boo  \citep{Boden05}
is based on the Infrared Flux method and yields: $R_{\rm
A}=2.474\pm0.096$~\rsol\ and  $R_{\rm B}=1.86\pm0.15$~\rsol. However,
a more precise and independent determination of the stellar radii
could be obtained, thanks to the small uncertainty on the masses
(0.3\%), from the knowledge of $\langle \Delta\nu\rangle$.
In particular for component B, the  measurement of
$\langle \Delta\nu\rangle$,  would  result in a significantly smaller
uncertainty on its location in the HR diagram  and, therefore, in
tighter constraints on the modelling. In Fig.~\ref{fig:radiusB} we
show the uncertainty on the radii deriving from $\langle
\Delta\nu\rangle$ values supposed to be known with an accuracy of  1
and 2 $\mu$Hz.

\begin{figure}
\begin{center}
\resizebox{.9\hsize}{!}{\includegraphics[]{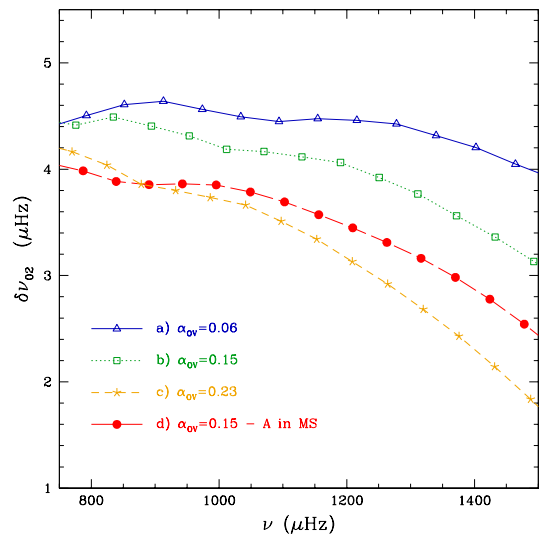}}
\caption{\small The small frequency separation $\delta\nu_{02}$ as a
function of frequency in models of component B corresponding to the
scenarios $a-d$ described in Fig.~\ref{fig:solution1}.}
\label{fig:sismoB}
\end{center}
\end{figure}

Information on the central structure of component B could be provided
by the small frequency separations $\delta\nu_{02}$. In
Fig.~\ref{fig:sismoB} we show  $\delta\nu_{02}$ as a function of the
frequency,
as expected for the models of 12~Boo~B corresponding to the four
scenarios of Fig.~\ref{fig:solution1}.
We notice that the dependence of $\delta\nu_{02}$ on $\nu$ is mainly
determined by the location of the discontinuity
in sound speed derivative  at the border of convective core, while the
age of the model changes the
mean value of $\delta\nu_{02}$.
The spread of $\delta\nu_{02}$ for the four considered scenarios
increases with the frequency
and is of the order of 2~$\mu$Hz at 1500~$\mu$Hz and of 1.5~$\mu$Hz at
$\nu_{max}$. 
The age difference between scenarios {\it b} and {\it d} (note that
both have $\alpha_{\rm OV}=0.15$)
implies  a shift of only 0.5$\mu$Hz in   $\delta\nu_{02}$.

Can $\delta\nu_{02}$ of component B help to discriminate between
scenarios {\it c} and {\it d}
in absence of avoided crossing detection?
Fig.~\ref{fig:sismoB} shows that the difference 
of $\delta\nu_{02}$ at $\nu_{\rm max}$ is only 0.2~$\mu$Hz and reaches
0.5~$\mu$Hz at 1500~$\mu$Hz.
Furthermore,  the frequency of $\ell=2$ modes will be affected by
rotational splitting (expected to be $\sim1-1.5\,\mu Hz$), so that an
accurate and reliable measurement
of the small separation (expected to be around 3 $\mu$Hz) will be a
rather difficult task.
Other frequency combinations, such as the small separation $d_{01}$
are known to be
very sensitive to the core structure of main sequence stars (see e.g.
\citealt{Roxburgh03} and references therein), but the larger number of
frequencies needed to compute these small differences,
increases even more their observational uncertainty.

\subsection{Probing the chemical composition gradient in the core}
\label{sec:profile}
Is there any other information that the knowledge of the frequency of an avoided crossing can give us?
The appearance of avoided crossings is due to the evolution of frequencies of gravity modes and their interaction  
with acoustic modes of similar frequency. 
The frequency spectrum of gravity modes, on its turn, is mainly determined by the behaviour of $N$ in  
the central regions of a star ($\nu \propto \int N/r\,dr$). 
Furthermore, the mean molecular weight gradient ($\nabla_\mu$) is determined by the evolutionary state, but
it can be also modified by different mixing   processes taking place in the radiative interior. 
If the  $N$ profile changes because of  a different   $\nabla \mu$, then we could 
 expect a signature in the {\it avoided crossings} frequencies.

\begin{figure*}
\begin{center}
{\includegraphics[angle=-90,width=0.8\textwidth]{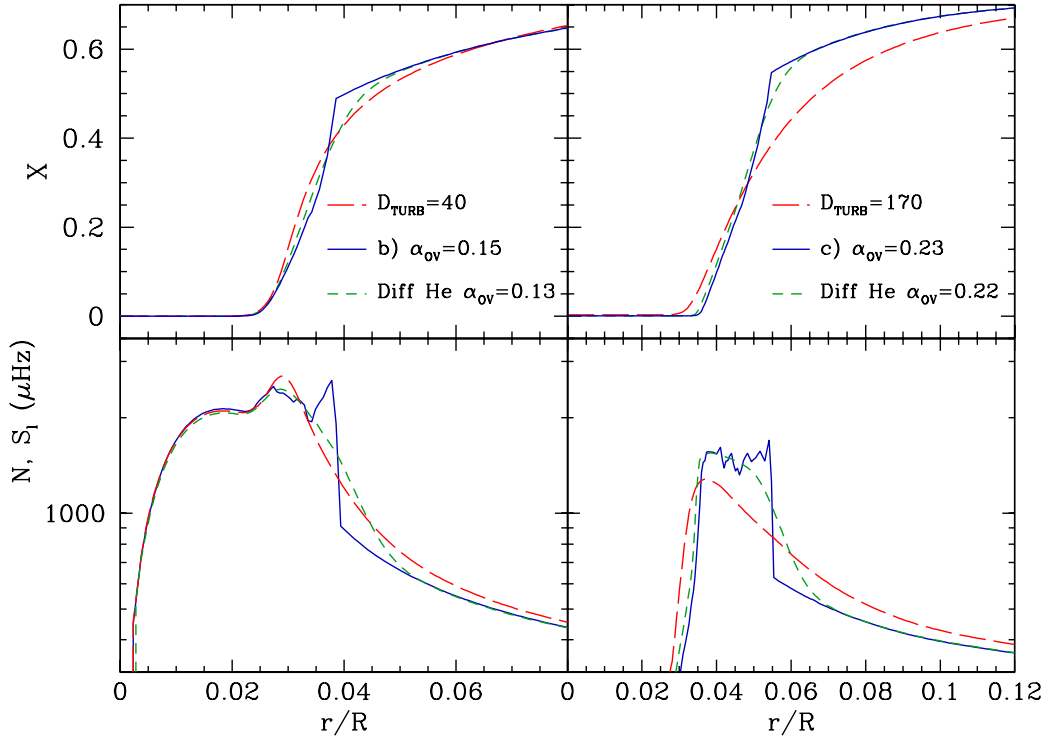}}
\caption{\small Hydrogen profile (\textit{upper panel}) and \BV frequency (\textit{lower panel}) in models computed with overshooting (continuous line), with turbulent mixing  (long-dashed line) and with diffusion of H and He (short-dashed lines)
}
\label{fig:core-dr}
\end{center}
\end{figure*}

\begin{figure}
\begin{center}
\resizebox{.9\hsize}{!}{\includegraphics[]{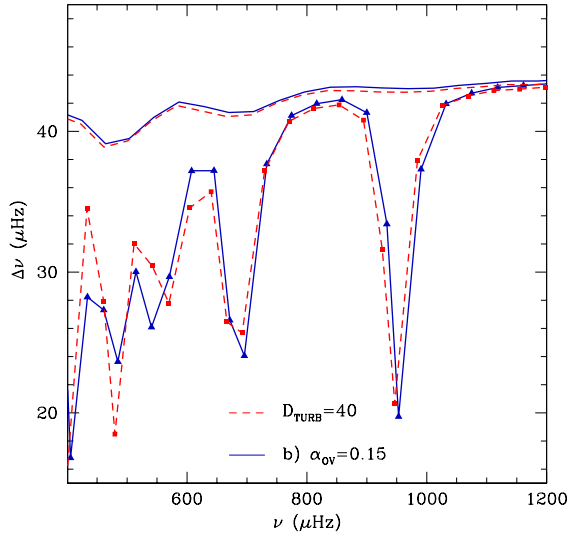}}
\caption{\small $\ell=0$ and $\ell=1$ large frequency separation in models of which chemical gradient and
\BV are plotted in Fig.~\ref{fig:core-dr}. Model A$b$ with overshooting $\alpha_{\rm OV}=0.15$ (continuous line), 
and model with turbulent mixing (long-dashed line). 
The different chemical composition gradient (see Fig.~\ref{fig:core-dr}) affects the frequencies of $\ell=1$ 
modes undergoing avoided crossings (lower curves) but not the frequencies of radial modes (upper curves).
}
\label{fig:largerot}
\end{center}
\end{figure}

In each of the evolutionary states {\it a} -- {\it d} we considered, the different mixing processes we 
analyse (classical and diffusive overshooting, microscopic and turbulent diffusion) provide 
models with a central structure showing a different $\nabla\mu$ and therefore different $N$ profiles. 
In Fig.~\ref{fig:core-dr}
we plot the hydrogen mass fraction  and $N$ profiles in the central region of 12~Boo~A models in the evolutionary states
{\it b}  and {\it c}  resulting from the calibrations with 
classical overshooting, with (short-dashed lines) and without microscopic diffusion (solid lines), 
and calibration with turbulent diffusion (long-dashed lines).
Though the evolutionary state is the same, the difference of chemical composition gradient 
(Fig.~\ref{fig:core-dr} upper panels) is clearly reflected on the  \BV frequency profile (Fig.~\ref{fig:core-dr} 
lower panels), and therefore on the frequency of g-modes and on the properties of the avoided crossings 
(see Fig.~\ref{fig:largerot}).
Consequently, the  determination of mixed-mode frequencies with a precision better than $\sim 2\,\mu$Hz for 
frequencies around 650~$\mu$Hz, would allow to determine
the properties of the chemical composition profile and, hence, of the physical mechanism responsible for the 
extra-mixing outside the core. The effect of different $\nabla\mu$  is even larger at lower
frequencies: at 500~$\mu$Hz, where the expected mode amplitude is slightly larger than the
half of the amplitude at $\nu_{\rm max}$, the difference between overshooting and $D_{\rm T}$ 
models reaches 5~$\mu$Hz.

We notice that such a valuable inference on the properties of the chemical composition gradient 
is made possible, in the example presented here above, by a trade-off between two effects. 
On one hand, the difference in $\nabla_{\mu}$,
which is due to a different mixing process, decreases as we consider more evolved models. 
On the other hand,
the number of g modes entering the domain of solar-like oscillations, and therefore the number of 
avoided crossings,  increases as the model evolves.
 This is the reason why, even though the differences between chemical composition profiles provided
by classical overshooting and by $D_{\rm T}$  models increase as we consider less evolved 
models (see Fig.~\ref{fig:core-dr}, lower panel), the 
lack of $\ell=1$ avoided crossing predicted in the domain of solar-like oscillations
for the scenarios {\it c} and {\it d}, 
makes difficult to infer information on the detailed properties of
the  $\mu$ gradient.

As mentioned above, models in the second gravitational contraction  present  $\ell=2$ avoided crossings in the 
solar-like spectrum domain. These $\ell=2$ modes are however less sensitive than $\ell=1$ modes to the $N$ profile,
and the difference between overshooting and microscopic diffusion profiles is not large  enough to be
reflected in $\ell=2$ frequency modes. Models in the scenario {\it c} including turbulent diffusion
show, however, very different $N$ in the central region, in particular, the model reaches the same location in 
the HR diagram with a smaller convective core, and this is reflected on a difference in the frequency of 
 modes undergoing an avoided crossing, of the order of 100~$\mu$Hz with respect to those in 
the overshooting models.

Concerning the main-sequence  calibrated models (both for  A and B components)
we note that while in overshooting models $\delta\nu_{02}$  as a function
of $\nu$ has a  slope  increasing with the value of the overshooting parameter,
for turbulent diffusion
models, $\delta\nu_{02}$ shows only a weak $\nu$-dependence.
That can be explained in terms of a different effect on the size of
the convective core.
While, in fact,  in overshooting models  the
increase of  $\alpha_{\rm OV}$ implies a larger core,  in turbulent
diffusion ones the increase
of  $D_{\rm T}$ parameter  decreases $\nabla\mu$ at the border of the
convective core, but leaves
the core  radius almost unaffected.
Besides,  the convective core for 12~Boo stellar masses is too small,
and the difference
between $\delta\nu_{02}$ slope for overshooting and $D_{\rm T}$ models
not large enough to be
detected in the  expected solar-like frequencies.

The effect of other diffusion processes such as the diffusive overshooting or the microscopic diffusion
on the $\mu$-profile  and on the frequencies of the avoided crossings are similar to those
described for the turbulent mixing.
The effect of microscopic diffusion is, however, not limited to the core: it also modifies the structure 
of the envelope by a decrease of the helium abundance in the outer convective zone and by the appearance 
of a chemical composition gradient  below the convective envelope. The combination of these effects leaves, 
as noticed  e.g. in \citet{Theado05}, a clear signature in the frequencies of acoustic modes. In order to 
fully exploit this seismic signature a high frequency resolution, that only long and uninterrupted 
observations can provide, is however needed.

\section{Conclusions}
\label{sec:conclusion}

We have presented a detailed modelling of the binary system 12~Bo\"otis by  fitting the
available observational constraints: effective temperatures, luminosities, chemical
composition, and high precision masses of both components.
As result of different 12~Boo  calibrations we found  a set of possible theoretical scenarios where 
the secondary is on the main sequence, whereas the primary has already left it.
Its precise evolutionary state, however,  may vary from the SOC (models A{\it c})
to a thick-shell-H-burning phase (models A{\it a}) by  varying the overshooting
parameter $\alpha_{\rm OV}$ from 0.23 to 0.06. These values slightly decrease (by 0.02) if 
microscopic diffusion is included in stellar modelling. 
Other central mixing processes, such as  diffusive overshooting (described by the $\beta$
parameter), and a turbulent mixing with a parametric turbulent diffusion coefficient ($D_{\rm T}$)
lead to the similar results. For each transport processes we found a possible range of values of 
the parameters $\alpha_{\rm OV}$, $\beta$ and $D_{\rm T}$ that place 12~Boo~B in the MS and 12~Boo~A
between the second gravitational contraction and the sub-giant phase.
The only way to place both components in the MS (models A{\it d}) is to assume a different 
efficiency of the mixing process in both components, for instance : $\alpha_{\rm OV,A}=0.37$, $\alpha_{\rm OV,B}=0.15$,
or $D_{\rm T,A}=330$~cm$^2$~s$^{-1}$, $D_{\rm T,B}=100$~cm$^2$~s$^{-1}$.
With the available observational constraints we are not able, however, to discriminate among the different 
scenarios and models proposed: it is cleat that additional and independent observational constraints are needed.

In Sec. \ref{sec:seismology} we have shown that the detection of the theoretically predicted 
solar-like oscillations in 12~Bo\"otis would provide a powerful test to discriminate between 
different scenarios. Indeed, the detection of avoided crossings
 in the primary oscillation spectrum  will allow a robust inference on the evolutionary state
  of the star and, therefore, on the amount of overshooting (or extra-mixing) from the core.
  The presence or absence  of $\ell=1$ mixed modes will allow  to discriminate between sub-giant and MS models,
  and the absence of both  $\ell=1$ and $\ell=2$ will allow to place 12~Boo~A in the MS, before the SOC.
      This possibility is especially exciting since it would imply
  that, even for stars with masses so close, the central mixing has might have been working with a significantly
  different efficiency, opening the door to important questions concerning transport processes.


As shown in Sec. \ref{sec:profile}, for a model in a given evolutionary state, a different chemical 
composition profile in the core can be due to a different physical process responsible for the extra-mixing. 
This affects the behaviour of the \BV frequency and hence the frequency of modes of mixed p- and g- character. 
Consequently, information on the mixing process taking place in the stellar center could be inferred from
the mixed mode frequencies. The precision required to reach this goal is, however, evolutionary state dependent.
While for a thin-shell-H-burning phase, a precision of 2~$\mu$Hz should be enough to  distinguish between 
the sharp chemical profile provided by overshooting and the smother one from a diffusive process, in
a slightly more evolved phase (thick-shell-H-burning phase) the effect would be  already erased.
On the other hand, as stellar age decreases, the number of mixed modes also decrease. As consequence, for
models in the second gravitational contraction, information on the mixing process can be derived from $\ell=2$ mixed modes 
only if the generated chemical profiles are quite different. 
For MS models, where no avoided crossings are predicted, the effect of the central mixing on 
the small separation $\delta\nu_{02}$ is not large enough to distinguish between different transport
processes.

Finally, the detection of solar-like oscillations also in the B component, will  reduce  the 
uncertainties on the HR location of 12~Boo~B, and
hence will significantly improve the observational constraints of the system.
Nevertheless, as all the models of component B are in the MS, no useful information concerning
the age of the system  will be add by the 12~Boo~B frequencies.

In conclusion,  we have shown that thanks to the precise knowledge of the masses of 12~Bo\"otis, 
and the additional constraints due to the binarity,  the search and detection  
of solar-like oscillations in  12~Boo,  hopefully by a spectroscopic multi-site campaign of observations, 
would mean an important step in the understanding of stellar evolution in the mass domain where the
the convective core appears.


\section*{Acknowledgements}
A.M and J.M acknowledge financial support from the Prodex-ESA Contract Prodex 8 COROT (C90199). CM thanks funding 
from COFIN 2004-Asteroseismology.

\bibliographystyle{mn2e}
\small
\bibliography{andrea}
\label{lastpage}
\end{document}